\documentclass[twocolumn,prl,floats,showpacs,superscriptaddress]{revtex4}

\usepackage{amsmath}
\usepackage{amssymb}
\usepackage{graphicx}
\usepackage[T1]{fontenc}
\usepackage{color}

\begin{document}

%\title{Crossover to a quasi-condensate in a weakly interacting trapped 1D Bose gas}
\title{Interaction-induced crossover versus finite-size condensation in a weakly interacting trapped 1D Bose gas}
\author{I. Bouchoule}
\affiliation{Laboratoire Charles Fabry, UMR 8501 du CNRS, 91 403 Orsay Cedex, France}
\author{K. V. Kheruntsyan}
\affiliation{ARC Centre of Excellence for Quantum-Atom Optics, School of Physical
Sciences, University of Queensland, Brisbane, Queensland 4072, Australia}
\author{G. V. Shlyapnikov}
\affiliation{Laboratoire de Physique Theorique et Modeles Statistiques, Universite
Paris-Sud, 91405 Orsay Cedex, France}
\affiliation{Van der Waals-Zeeman Institute, University of Amsterdam, 1018 XE Amsterdam, The Netherlands}
\date{\today }

\begin{abstract}
We discuss the transition from a fully 
decoherent to a (quasi-)condensate regime in a
harmonically trapped weakly interacting 1D Bose gas.
By using analytic approaches and verifying them against exact numerical solutions, we find a characteristic crossover temperature and crossover atom number that depend 
on the interaction strength and the trap frequency. We then identify the conditions for observing either an \textit{interaction-induced} crossover 
scenario or else a \textit{finite-size} Bose-Einstein condensation phenomenon 
characteristic of an \textit{ideal} trapped 1D gas. 
\end{abstract}

\maketitle

One-dimensional (1D) Bose gases are remarkably rich physical 
systems exhibiting properties not encountered in 2D or 
3D \cite{Girardeau1960,Lieb-Liniger,Yang}. Here we study the 1D
model of bosons interacting via a repulsive delta-function 
potential, which plays a fundamentally important role in quantum 
many-body physics. The reason is that the model is exactly 
solvable \cite{Lieb-Liniger,Yang} and it is now
experimentally realizable with ultracold alkali atoms in highly 
anisotropic trapping potentials 
(see Ref. \cite{review1D} for a review). This means there are unique opportunities 
for accurate tests of theory that were previously unavailable, in turn 
leading to the development of fundamental knowledge of interacting 
many-body systems in low dimensions.

In this paper, we analyze the properties of the 1D Bose gas in the weakly
interacting regime, where the dimensionless interaction parameter $\gamma
=mg/(n\hbar ^{2})$ is small, $n$ being the linear density, $%
m $ the atom mass, and $g$ the 1D coupling constant. This is opposite to
Girardeau's regime of \textquotedblleft
fermionization\textquotedblright\ \cite{Girardeau1960}
achieved in the limit of strong interactions and 
subject of many recent studies \cite{1D-experiments}.
Our motivation for the study of the weakly interacting regime is to reveal the 
nature of the transition to a Bose-condensed state in a harmonically 
trapped system.

For a uniform weakly interacting 1D Bose gas, one has a smooth
interaction-induced crossover  to a quasi-condensate which is a 
Bose-condensed state with a fluctuating phase. This occurs when the temperature $%
T $ becomes smaller than $T_{d}\sqrt{\gamma }$ \cite%
{Kheru2003,Kheru2005,Castin1Dclass,Quasibec-Castin}, where $T_{d}=\hbar
^{2}n^{2}/2m$ is the temperature of quantum degeneracy (in energy units, $%
k_{B}=1$). For a harmonically trapped 1D gas with weak interactions a
similar crossover scenario is expected \cite{Kheru2005}. However,
due to the presence of the trapping potential the interaction-induced crossover enters into a
competition with Bose-Einstein condensation (BEC) predicted to occur in the
ideal gas limit \cite{Ketterlecond1D} as a macroscopic occupation of the
ground state. For a given atom
number $N$, this condensation phenomenon occurs at temperature $T_{C}\simeq
N\hbar \omega /ln(2N)$. It is a purely finite-size effect and disappears
in the thermodynamic limit \cite{Bagn91} where $N$ tends to infinity while
the peak density $n_{0}$ is kept constant (this implies that the trap
oscillation frequency $\omega $ tends to zero in such a way that $N\omega =%
\mathrm{const}$). The interaction-induced crossover to a quasi-condensate, 
on the other hand, persists in the thermodynamic limit.

Thus, for sufficiently weak confinement one expects to observe an
interaction-induced crossover to a quasi-condensate, rather than a
finite-size BEC. The situation is reversed for strong confinement. 
Here, we
identify the parameters of the interaction-induced crossover and find the 
conditions
that enable 
the realization of either of these two competing scenarios.

We start by briefly reviewing the physics of a uniform 1D Bose gas in the
thermodynamic limit, in the case of very weak interactions $\gamma \ll 1$.
For $T\ll T_{d}\sqrt{\gamma }$, the gas is in the quasi-condensate 
(Gross-Pitaevskii) regime where 
the density fluctuations are suppressed and the gas is coherent on a 
distance scale smaller than the phase coherence length:~
Glauber's local pair correlation function is
reduced below the ideal gas level of 2 and is close to 1 \cite%
{Kheru2003,Kheru2005,Castin1Dclass,Quasibec-Castin}. In this regime the
chemical potential is positive and well approximated by $\mu
\simeq gn$. 
For $T\gg T_{d}\sqrt{\gamma }$, the gas is in the fully decoherent regime:
interactions between the atoms have a small effect on the equation of state
and the local pair correlation is close to that of an ideal Bose gas \cite%
{Kheru2003}. This regime contains the quantum decoherent domain 
$T_d\sqrt{\gamma}\ll T \ll T_d$.
 In the decoherent regime, the chemical potential $\mu $ is
negative and the equation of state is well approximated by that of the ideal
Bose gas: 
\begin{equation}
n=\int_{-\infty }^{\infty }\frac{dk/(2\pi )}{e^{(\hbar ^{2}k^{2}/2m-\mu
)/T}-1}=\sqrt{\frac{mT}{2\pi \hbar ^{2}}}\sum_{j=1}^{\infty }\frac{e^{j\mu
/T}}{j^{1/2}}.  \label{eq.muid}
\end{equation}

The crossover between the decoherent and the quasi-condensate regimes ($%
T\sim T_{d}\sqrt{\gamma }$) corresponds to the density of the order of $n_{%
\mathrm{co}}=(mT^{2}/\hbar ^{2}g)^{1/3}$. Using the crossover density $n_{%
\mathrm{co}}$ is convenient for analyzing the properties of the gas at a
constant temperature and varying $n$. In this sense, the quantum decoherent
regime corresponds to $n_{d}\ll n\ll t^{1/6}n_{d}\simeq n_{\mathrm{co}}$,
where $t={T}/{T_{d}\gamma ^{2}}=2\hbar ^{2}T/mg^{2}$ is a dimensionless
temperature parameter which is independent of the density and is large, and
$n_{d}=\sqrt{mT}/\hbar $ is the density of quantum degeneracy at a given $T$%
. The width of the quantum decoherent region in terms of densities increases
with $t$.

\begin{figure}[tbp]
%\vspace{4.2cm} 
\includegraphics[height=3.90cm]{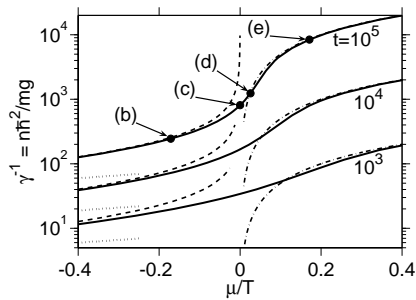}
\caption{Equation of state of the uniform weakly-interacting 1D Bose gas for
three different values of the temperature parameter $t=2T\hbar ^{2}/mg^{2}$.
The exact numerical result (solid line) is compared with the behavior in the
quasi-condensate regime (dash-dotted lines) and with the ideal Bose gas
result of Eq.~(\protect\ref{eq.muid}) (dashed lines). The straight dotted
lines correspond to the classical (Boltzmann) ideal gas. }
\label{fig.nvsmu}
\end{figure}

In Fig.~\ref{fig.nvsmu} we illustrate the properties of the weakly
interacting uniform gas by plotting the linear density as a function of the
chemical potential for three different values of the temperature parameter $%
t $. The exact numerical results \cite{Kheru2003} based on the
finite-temperature solution \cite{Yang} to the Lieb-Liniger model \cite%
{Lieb-Liniger} are compared with both the ideal Bose gas equation of state (%
\ref{eq.muid}) in the region of $\mu <0$ and with the quasi-condensate
equation of state corresponding to $\mu \simeq gn>0$. For a given
temperature, the crossover from the decoherent regime to the
quasi-condensate corresponds to $\mu $ going from negative to positive. We
obtain $n(\mu =0,T)\simeq 0.6n_{\mathrm{co}}$ within 20\% accuracy as long as $%
t>10^{3}$. Note that the values of $t$ as high as $10^{3}$ are required to
ensure that the gas is highly degenerate at the crossover.

We now turn to the analysis of a harmonically trapped 1D gas and find the
crossover temperature $T_{\mathrm{co}}$ and crossover atom number 
$N_{\mathrm{co}}$ around which the gas enters the quasi-condensate
regime. For small trap frequencies $\omega $, the density profile of the 
gas can be described using the local
density approximation (LDA) \cite{Kheru2005}. In this treatment, the 1D
density $n(z)$ as a function of the distance $z$ from the trap centre is
calculated using the uniform gas equation of state in which the chemical
potential $\mu $ is replaced by its local value $\mu (z)=\mu _{0}-m\omega
^{2}z^{2}/2$, where $\mu _{0}$ is the global chemical potential. Within the
LDA, the uniform results remain relevant and imply, in particular,
that the gas enters the quasi-condensate regime in the trap centre once $\mu
_{0}$ changes sign. In addition, as long as the peak density $n_{0}=n(0)$
fulfills the condition $n_{0}\ll n_{\mathrm{co}}$ the entire gas is in the
decoherent regime and the equation of state is well approximated by Eq.~(\ref%
{eq.muid}) in which $n$ and $\mu $ are replaced by $n(z)$ and $\mu (z)$.
Integrating $n(z)$ over $z$ and taking the sum over $j$ 
gives a relation between the total atom number and $\mu_0$:
\begin{equation}
N=-{T}/{(\hbar \omega )}\ln (1-e^{\mu _{0}/T}),\;\;(\mu _{0}<0).
\label{eq.Ntot}
\end{equation}

\begin{figure*}[tbp]
\centerline{\includegraphics[height=3.1cm]{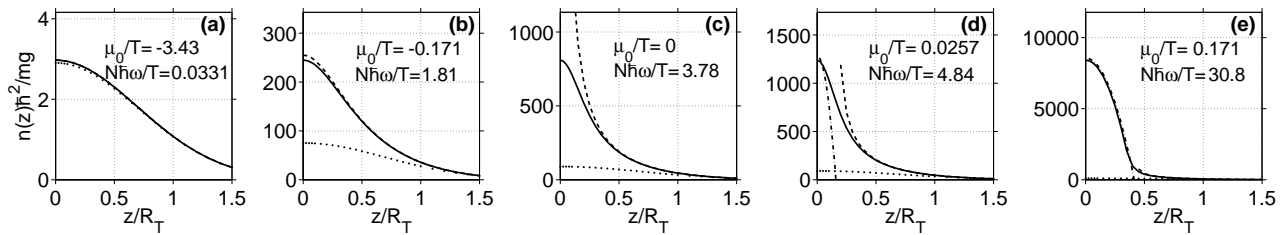}}
\caption{Density profiles of a 1D Bose gas in a harmonic trapping potential
for five different values of the ratio $\protect\mu _{0}/T$ and a fixed
value of the temperature parameter $t=2\hbar ^{2}T/mg^{2}=10^{5}$. The exact
numerical solution (solid line) is compared with the ideal Bose gas
distribution (dashed line), classical Boltzmann distribution (dotted line),
and Thomas-Fermi distribution in the Gross-Pitaevskii regime (dashed-dotted
line). The resulting values of the dimensionless ratio $N\hbar\omega/T$, 
following the exact solutions, are also shown. 
The distance from the trap center $z$ is in units of $%
R_{T}=(2T/m\protect\omega ^{2})^{1/2}$. All calculations are done within the
LDA using the equation of state for the homogeneous gas shown in Fig. 1, with
$\protect\mu _{0}$ and $n(0)$ in (b)-(e) being the same as $\protect\mu $ and 
$n$ indicated by the points (b)-(e) in Fig.~\ref{fig.nvsmu}.}
\label{fig.profiles}
\end{figure*}

As mentioned above, for very large values of $t$ the crossover to the
quasi-condensate occurs under conditions where the gas is highly degenerate
in the centre, with $n_{0}\gg n_{d}=\sqrt{mT}/\hbar $. Assuming that this is
the case and taking into account that the degeneracy condition is equivalent
to $|\mu _{0}|/T\ll 1$, Eq.~(\ref{eq.Ntot}) can be rewritten as%
\begin{equation}
N\simeq {T}/{(\hbar \omega )}\ln \left( {T}/{|\mu _{0}|}\right) .  \label{N}
\end{equation}%
Under these conditions, as Eq.~(\ref{eq.muid}) reduces to $n\simeq \sqrt{%
mT^{2}/2\hbar ^{2}|\mu |}$ for $|\mu |\ll T$, the density
profile  develops a sharp central peak which is well approximated by 
%\cite{comment3}
\begin{equation}
n(z)\simeq \sqrt{\frac{mT^{2}}{2\hbar ^{2}}}\frac{1}{\sqrt{|\mu _{0}|+m\omega ^{2}z^{2}/2}},
\label{eq.peakprofile}
\end{equation}
and extends up to distances $|z|\lesssim R_{T}=\sqrt{2T/m\omega ^{2}}$.

The analysis made above is valid as long as $n_{0}\ll n_{\mathrm{co}}$. Using Eq.
(\ref{eq.peakprofile}) and the expression for $n_{\mathrm{co}}$, the
condition $n_{0}\ll n_{\mathrm{co}}$ can be rewritten as 
\begin{equation}
|\mu _{0}|\gg m^{1/3}(gT/\hbar )^{2/3}.  \label{eq.mudec}
\end{equation}

Using Eq.~(\ref{N}) to relate $\mu _{0}$ to the total atom number,
Eq.~(\ref{eq.mudec}) leads
to the condition that the gas is in the decoherent regime as long as $N\ll
N_{\mathrm{co}}$, where 
\begin{equation}
N_{\mathrm{co}}\simeq {T}/({\hbar \omega })\ln \left( \hbar
^{2}T/mg^{2}\right) ^{1/3}=(T/3\hbar \omega )\ln (t/2)  \label{eq.Nco}
\end{equation}%
is the characteristic atom number at the crossover.
As we mentioned earlier, one should have $t\gg 10^{3}$ for obtaining a
highly degenerate gas at the crossover. Under this condition, Eq.~(\ref%
{eq.Nco}) can be approximately inverted to yield, for a given $N$, a
crossover temperature 
\begin{equation}
T_{\mathrm{co}}\simeq {N\hbar \omega }/{\ln \left( N\hbar ^{3}\omega
/mg^{2}\right) ^{1/3}}.  \label{eq.Tco}
\end{equation}

We emphasize that our results are obtained within the LDA which is
valid if the characteristic correlation length $l_{c}$ of
density-density fluctuations is much smaller than the typical length scale 
$L$ of density variations. The correlation length is 
$l_{c}\simeq \hbar /\sqrt{m|\mu _{0}|}$ in both the quantum
decoherent and quasi-condensate regimes \cite{Kheru2003,Kheru2005}.
Approaching the crossover
from the decoherent regime we replace $|\mu _{0}|$ by the rhs of
Eq.~(\ref{eq.mudec}), while approaching it from the
quasi-condensate regime we use $\mu _{0}\simeq gn_{\mathrm{co}}$. In both
cases, one obtains $l_{c}\simeq \hbar ^{4/3}/(m^{2}gT_{\mathrm{co}})^{1/3}$.
The length scale $L$ can be estimated as the distance from the trap center
where the density is halved compared to the peak density $n_{0}$. Approaching
the crossover from the decoherent side, Eq.~(\ref%
{eq.peakprofile}) gives $L\simeq \sqrt{|\mu
_{0}|/m\omega ^{2}}\simeq (gT_{\mathrm{co}}/m\hbar \omega ^{3})^{1/3}$. On 
 the quasi-condensate side, we use the Thomas-Fermi parabola and obtain $L\simeq 
\sqrt{2n_{\mathrm{co}}g/m\omega ^{2}}$, which gives approximately the same
result. One then easily sees that the condition of validity of the LDA, $%
l_{c}\ll L$, is reduced to
\begin{equation}
\omega\ll\omega _{\mathrm{co}}\equiv (mg^{2}T^{2}/\hbar ^{5})^{1/3}.  
\label{eq.condco}
\end{equation}%

If this inequality is not satisfied then the LDA breaks down and one has to
take into account the discrete structure of the trap 
energy levels. In this case, analytic approaches incorporating both the finite-size 
effects and small but finite interaction strength 
are absent in the vicinity of the transition to a quasi-condensate, and we adopt the 
ideal gas treatment of Ref. \cite{Ketterlecond1D}. 
For a fixed temperature, 
this treatment predicts a finite-size BEC 
at a critical atom number 
$N_{C}=T/(\hbar \omega) \ln (2T/\hbar\omega)$. 
It is clear that the finite-size BEC phenomenon will 
prevail the interaction-induced 
crossover scenario if $N_{C}<N_{\mathrm{co}}$. 
In fact, the opposite inequality, $N_{C}>N_{\mathrm{co}}$,
is equivalent to that of Eq.~(\ref{eq.condco}), which makes our 
analysis self-consistent and implies that the condition 
of validity of the LDA, $\omega\ll\omega _{\mathrm{co}}$, serves as the simultaneous 
criterion for observing the interaction-induced crossover, while the opposite 
condition corresponds to finite-size condensation. 
At a constant $N$, the criterion for observing the interaction-induced crossover
can be obtained from Eq.~(\ref{eq.condco}) by replacing $T$ with $N\hbar\omega/\ln(2N)$.
The opposite criterion leading to the finite-size BEC has been previously 
found in Ref. \cite{Phasefluctu-Petrov} from the condition $gn_{0}\ll \hbar \omega$.

In the following, we analyze the properties of the interaction-induced crossover, 
subject to inequality (\ref{eq.condco}). 
Since $t\gg 10^{3}$ in the
regime of interest, Eq.~(\ref{eq.Nco}) written as $T_{\mathrm{co}}=3N\hbar
\omega /\ln (t_{\mathrm{co}}/2)$ shows that the crossover temperature 
is lower than the characteristic temperature of
quantum degeneracy of a harmonically trapped gas $N\hbar \omega $. Thus, $T_{%
\mathrm{co}}$ represents a more accurate and lower estimate of the crossover
temperature to the quasi-condensate regime compared to the inequality $T\ll N\hbar\omega$ 
given in Ref. \cite{Phasefluctu-Petrov}. For extremely large values of $t$, 
the present treatment
identifies an intermediate temperature interval $T_{\mathrm{co}}\ll T \ll
N\hbar \omega $ which accommodates the decoherent quantum regime. Here the
gas is degenerate and is well described within the ideal Bose gas approach.

\begin{figure}[b]
%\vspace{4.3cm} 
\includegraphics[height=4.3cm]{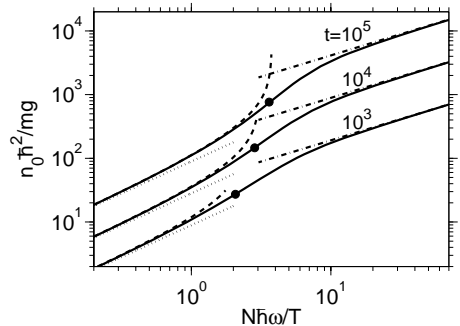}
\caption{Peak density $n_{0}$ (in units of $mg/\hbar ^{2}$) of a trapped gas
versus $N\hbar \protect\omega /T$ for three values of $t=2\hbar ^{2}T/mg^{2}$%
. The three black dots show the respective crossover values of $N_{\mathrm{co%
}}\hbar \protect\omega /T$ from Eq.~(\ref{eq.Nco}). The different
lines are as in Fig.~\ref{fig.nvsmu}.}
\label{fig.peakdensity}
\end{figure}

Fig.~\ref{fig.profiles} shows density profiles for
different values of the chemical potential at a fixed temperature parameter
$t=2\hbar ^{2}T/mg^{2}=10^{5}$. The graph (e) corresponds to the
quasi-condensate regime. The graph (c) shows the density profile at the
crossover, and we find that the corresponding atom number, $N\simeq
3.78T/\hbar \omega $, is in good agreement with the value $N_{\mathrm{co}%
}\simeq 3.61T/\hbar \omega $ predicted by Eq.~(\ref{eq.Nco}). The decoherent
regime is clearly seen in graphs (a) and (b). Although the inequality 
$T_{\mathrm{co}}\ll N\hbar\omega$ is barely satisfied, the features of  
the quantum decoherent regime are seen in (b): the density
profile is described to better than 10\% by the ideal Bose gas
approach and differs strongly from the classical Boltzmann distribution.

To provide a better connection with experimentally measurable quantities we
plot in Fig.~\ref{fig.peakdensity} the peak density $n_{0}$ versus $N\hbar
\omega /T$ for three different values of the temperature parameter $t$. In
all cases we give the comparison with the classical Boltzmann gas, the ideal
Bose gas, and the quasi-condensate predictions. The ideal Bose gas
prediction connects the Boltzmann behavior $n_{0}=N\omega \sqrt{m/2\pi T}$
to the degenerate behavior $n_{0}=(\sqrt{mT}/\hbar )\exp (N\hbar \omega /2T)$,
whereas in the quasi-condensate regime $n_{0}$ scales as $\propto N^{2/3}$. 
The scaling of the peak density $n_{0}$ as a function of $N$ and the
sequence of changes between power laws and an exponential can serve as a
signature of the transitions between different regimes. 
This includes the quantum decoherent regime, which
becomes more pronounced when increasing the parameter $t$ and is 
already seen for $t=10^{5}$.

The sufficient condition for realizing the 1D regime in a harmonically
trapped, weakly interacting gas is $T\ll \hbar \omega _{\perp }$, where $%
\omega _{\perp }$ is the transverse oscillation frequency. If the oscillator
length $l_{\perp }=\sqrt{\hbar /m\omega _{\perp }}$ is much larger than the
3D scattering length $a$, the 1D coupling is given by $%
g\simeq 2\hbar ^{2}a/ml_{\perp }^{2}$ \cite{Olshanii}. 
The condition for the interaction-induced crossover, $\omega
\ll \omega_{\mathrm{co}}$, can then be rewritten as 
\begin{equation}
\omega \ll \omega _{\perp }(T/\hbar \omega _{\perp })^{2/3}(a/l_{\perp
})^{2/3}.  \label{eq.condco1D}
\end{equation}
Taking $\omega _{\perp }/2\pi $ in the range from 1 to 30 kHz and 
$T\simeq 0.2\hbar \omega _{\perp }$ ($T$ is ranging from 10 to 300
nK), one can see that for most of the alkali atoms with typical scattering
lengths in the range of few nanometers, the inequality (\ref{eq.condco1D}) 
is well satisfied with $\omega$ of a few Hertz commonly used in practice.
Thus, the conditions for realizing the interaction-induced crossover are 
relatively easy to satisfy, unless the scattering length is extremely 
small ($a<0.1$ nm). On the other hand, the condition to observe the 
quantum decoherent regime before the interaction-induced crossover is more 
demanding as it requires, in addition to Eq.~(\ref{eq.condco1D}), a very 
large value of the parameter $t$. Rewriting the 1D 
inequality $T\ll \hbar \omega _{\perp }$ 
as $a\ll l_{\perp }/\sqrt{2t}$ we immediately see that even at 
$t=10^{5}$, where one only starts to see the features of this regime,
one needs to use light atoms (large $l_{\perp }$) and/or a very small 
scattering length in order to satisfy $a\ll 2\times 10^{-3}l_{\perp }$. 

A favorable system for fulfilling these conditions is a 1D gas of $^{7}$Li 
atoms in the $F=1,m=-1$ hyperfine state, where the scattering length can be
tuned from very large to extremely small values using an 
open-channel dominated Feshbach resonance \cite{theseFlorian}. By taking, for example, 
$\omega/2\pi  \simeq 4$ Hz, $\omega_{\perp}/2\pi \simeq 4$ kHz, 
$T\simeq 0.2\hbar \omega_{\perp }$ (40~nK), and varying $a$ 
from $20$ to $0.2$ nm one can increase $t$ from $60$ to 
$6\times 10^{5}$ and see how a direct interaction-induced crossover 
from a classical gas to a quasi-condensate regime transforms to
accommodate the intermediate quantum decoherent regime. 
The same system can also be used to observe the finite-size BEC
scenario, which requires the inequality opposite 
to Eq.~(\ref{eq.condco1D}) and hence a reduction of the scattering length to $a\simeq 0.01$ nm.

In conclusion, we have identified the conditions for realizing either 
a finite-size BEC phenomenon or an interaction-induced crossover 
to a coherent, quasi-condensate state in a harmonically trapped 1D Bose gas.
In the later case, we distinguish between a direct crossover from the 
classical decoherent regime and a crossover through the intermediate 
quantum decoherent regime.
Furthermore, one can expect that the physics of the interaction-induced 
crossover remains approximately valid for $T\sim\hbar \omega _{\perp }$, where 
the gas is no longer in the 1D regime but is near the 3D-1D boundary. 
This conjecture is supported by the results of
recent experiments \cite{exp1,exp2}. In Ref. \cite{exp1} a gas at $T\simeq
2\hbar \omega _{\perp }$ was produced with a density profile well described
within a degenerate ideal gas approach. This means that the crossover to a
quasi-condensate was likely to involve the features of the decoherent
quantum regime. Finally, we note that the interaction-induced crossover 
through a well pronounced decoherent quantum regime 
would be easier to produce in a quartic or box-like potential \cite{Raizen}. 

The authors acknowledge stimulating discussions with P. Drummond, D.
Gangardt, and C. Salomon. This work was supported by the IFRAF Institute. 
KK acknowledges support from the ARC. IB and GS acknowledge support from 
the ANR (grants NT05-2-42103 and 05-Nano-008-02). GS also acknowledges 
the Netherlands's FOM program on quantum gases. LPTMS is a mixed research 
unit no. 8626 of CNRS and Universite Paris-Sud.


\begin{thebibliography}{99}
\bibitem{Girardeau1960} M. D. Girardeau, J. Math. Phys. (N.Y.) \textbf{1}, 516
(1960).


\bibitem{Lieb-Liniger} E. H. Lieb and W. Liniger, Phys. Rev. \textbf{130},
1605 (1963).

\bibitem{Yang} C. N. Yang and C. P. Yang, J. Math. Phys. \textbf{10}, 1115
(1969).

\bibitem{review1D} D. S. Petrov, D. M. Gangardt, and G. V. Shlyapnikov, J.
Phys. IV (France) \textbf{116}, 5 (2004); Y. Castin, \textit{ibid.} \textbf{%
116} 89 (2004).

\bibitem{1D-experiments} H. Moritz \textit{et al.}, Phys. Rev. Lett. \textbf{91},
250402 (2003); B. L. Tolra \textit{et al.}, \textit{ibid.} 
\textbf{92}, 190401 (2004); T. Kinoshita, T. Wenger, and D. S. Weiss, \textit{ibid.} 
\textbf{95}, 190406 (2005); B. Paredes \textit{et al.}, Nature \textbf{429}, 277 (2004).

\bibitem{Kheru2003} K. V. Kheruntsyan \textit{et al.}, Phys. Rev. Lett. 
\textbf{91}, 040403 (2003).

\bibitem{Kheru2005} K. V. Kheruntsyan \textit{et al.}, Phys. Rev. A \textbf{%
71}, 053615 (2005).

\bibitem{Castin1Dclass} Y. Castin \textit{et al.}, J. Modern Opt. \textbf{47}%
, 2671 (2000).

\bibitem{Quasibec-Castin} C. Mora and Y. Castin, Phys. Rev. A \textbf{67},
053615 (2003).

%\bibitem{comment1} While certain features of the crossover were discussed in
%Ref. \cite{Kheru2005}, no explicit parameters were identified for the
%crossover temperature or crossover atom number.

\bibitem{Ketterlecond1D} W. Ketterle and N. J. van Druten, Phys. Rev. A 
\textbf{54}, 656 (1996).

\bibitem{Bagn91} V. Bagnato and D. Kleppner, Phys. Rev. A \textbf{44}, 7439
(1991).

\bibitem{Phasefluctu-Petrov} D. Petrov, G. Shlyapnikov, and J. Walraven,
Phys. Rev. Lett. \textbf{85}, 3745 (2000).

%% \bibitem{comment3} 
%% If $\mu/T\ll1$, many terms  contribute to the sum in Eq.\ref{eq.muid}  
%% and the sum can be approximated by an integral. This gives 
%% $n\simeq \sqrt{mT^{2}/(2\hbar^{2}|\mu |)}$.

\bibitem{Olshanii} M. Olshanii, Phys. Rev. Lett. \textbf{81}, 938 (1998).

\bibitem{theseFlorian} F. Schreck, Annales de Physique \textbf{28}, 1 (2003);
L. Khaykovich \textit{et al.}, Science \textbf{296}, 1290 (2002); V. A. Yurovski, Phys. Rev. A \textbf{73}, 052709 (2006).

%\bibitem{Olshanii2} T. Bergeman, M. G. Moore, and M. Olshanii, Phys. Rev.
%Lett. \textbf{91}, 163201 (2003).

%\bibitem{Leduc} S. Moal \textit{et al.}, cond-mat/0509286.

\bibitem{exp1} J. Esteve \textit{et al.}, Phys. Rev. Lett. \textbf{96},
130403 (2006).

\bibitem{exp2} J. Trebbia \textit{et al.}, quant-ph/0606247; here, 
the observed quasi-condensate formation could not be
explained by a BEC-like transition within the Hartree-Fock
theory.

\bibitem{Raizen} T. P. Meyrath \textit{et al.}, Phys. Rev. A \textbf{71},
041604(R) (2005).

\end{thebibliography}
\end{document}